\documentstyle[prl,aps]{revtex} 
\tighten
\begin{document}
\draft
\title{Universality in two-dimensional Kardar-Parisi-Zhang growth}
\author{F. D. A. Aar\~ao Reis}
\address{
Instituto de F\'\i sica, Universidade Federal Fluminense,\\
Avenida Litor\^anea s/n, 24210-340 Niter\'oi RJ, Brazil}
\date{\today}
\maketitle

\begin{abstract}
We analyze simulations results of a model proposed for etching of a
crystalline solid and results of other discrete models in the $2+1$-dimensional
Kardar-Parisi-Zhang (KPZ) class. In the steady states, the moments $W_n$ of
orders $n=2,3,4$ of the heights distribution are estimated. Results for the
etching model, the ballistic deposition (BD) model and the
temperature-dependent body-centered restricted solid-on-solid model (BCSOS)
suggest the universality of the absolute value of the
skewness $S \equiv {W_3} /{{W_2}^{3/2}}$ and of the value of the kurtosis
$Q \equiv {W_4} / {{W_2}^{2}} -3$. The sign of the skewness is the same of
the parameter $\lambda$ of the KPZ equation which represents the process in the
continuum limit. The best numerical estimates, obtained from the etching
model, are $|S| = 0.26\pm 0.01$ and $Q=0.134\pm 0.015$. For this model,
the roughness exponent $\alpha = 0.383\pm 0.008$ is obtained, accounting for a constant correction term (intrinsic width) in the scaling of the squared
interface width. This value is slightly below previous estimates of extensive
simulations and rules out the proposal of the exact value $\alpha=2/5$. The
conclusion is supported by results for the ballistic deposition model.
Independent estimates of the dynamical exponent and of the growth exponent are
$1.605\leq z\leq 1.64$ and $\beta = 0.229\pm 0.005$, respectively, which are
consistent with the relations $\alpha+z=2$ and $z=\alpha /\beta$.
\end{abstract}
\date{\today}


\section{Introduction}
\label{intro}

Surface growth processes and deposition of thin films have attracted much
interest from the technological point of view~\cite{frontiers,barabasi,krug}
and motivated the proposal of continuum and discrete models for surface and
interface growth, which also play an important role in the field of
non-equilibrium Statistical Mechanics. One of the most important
phenomenological theories is that of Kardar, Parisi and Zhang
(KPZ)~\cite{kpz}, in which the time evolution of the interface described by the
height function $h$ at position $\vec{x}$ and time $t$ is given by the KPZ
equation
\begin{equation}
{{\partial h}\over{\partial t}} = \nu{\nabla}^2 h + {\lambda\over 2}
{\left( \nabla h\right) }^2 + \eta (\vec{x},t) .
\label{kpz}
\end{equation}
Here $\nu$ represents a surface tension,
$\lambda$ represents the excess velocity and $\eta$ is a Gaussian
noise~\cite{barabasi,kpz} with zero mean and variance $\langle
\eta\left(\vec{x},t\right) \eta(\vec{x'},t') \rangle = D\delta^d
(\vec{x}-\vec{x'}) \delta\left( t-t'\right)$, where $d$ is the
dimension of the substrate.

The interface width $\xi(L,t) = {\left< \overline{h^2} - \overline{h}^2
\right> }^{1/2}$ characterizes the roughness of the interface,
for growth in a substrate of length $L$ (overbars 
denote spatial averages and angular brackets denote configurational
averages). For short times, the interface width scales as
\begin{equation}
\xi\sim t^{\beta} ,
\label{defbeta}
\end{equation}
with $\beta$ called the roughness exponent.
For long times, a steady state is attained and the width saturates at
\begin{equation}
\xi_{sat}\sim L^{\alpha} ,
\label{defalpha}
\end{equation}
with $\alpha$ called the roughness exponent. Eqs. (\ref{defbeta}) and
(\ref{defalpha}) correspond to limits of the dynamical scaling relation of
Family and Vicsek~\cite{famvic}
\begin{equation}
\xi \approx L^{\alpha} f\left( tL^{-z}\right) ,
\label{fv}
\end{equation}
where the dynamical exponent $z=\alpha /\beta$
characterizes the crossover from the growth regime to the steady state. 
Galilean invariance gives $\alpha+z=2$ for KPZ in all
dimensions~\cite{barabasi,krug}. The exact scaling exponents
are known in $d=1$, but no exact value was already obtained in two or more
dimensions~\cite{barabasi,krug}.

Many discrete models fall into the KPZ class, such as the
restricted solid-on-solid (RSOS) model of Kim and Kosterlitz~\cite{kk} and
ballistic deposition (BD)~\cite{vold}. Numerical estimates of the
scaling exponents in $d=1$ are consistent with the exact
values~\cite{kk,ball,balfdaar} and simulations in $d=2$ are frequently used to
estimate them. Most of the reported values of $\alpha$ range from $\alpha=0.37$
to $\alpha=0.4$~\cite{kk,devillard,liu,forrest,alanissila}, which is confirmed
by numerical solutions of the KPZ equation~\cite{amar,moser,beccaria,giada}.

In 1998, assuming certain properties of the height
correlation functions, L\"assig obtained a quantization condition for the KPZ
exponents which gave $\alpha = 2/5$ as the only solution consistent with the
range of numerical estimates~\cite{lassig}. Another consequence of
his work was that the moments of the steady state height distribution,
\begin{equation}
W_n \equiv {\left< \overline{ {\left( h-\overline{h}\right) }^n } \right> } ,
\label{defmoments}
\end{equation}
would obey power-counting, i. e. they would scale as $W_n \sim L^{n\alpha}$.
The second moment is the squared interface width $\xi^2$.
Moreover, the validity of his assumptions requires that the steady state
distribution is skewed (non-zero third moment), contrary to
the one-dimensional case (Gaussian distribution).

Recent numerical results of Chin and den Nijs for the RSOS and the
body-centered solid-on-solid (BCSOS) models were consistent with
$\alpha=0.4$~\cite{chin}, but extensive simulations of Marinari et al
for the RSOS model rule out that value~\cite{marinari}. A recent
study of the KPZ equation in the mode-coupling approximation provided an
estimate $z\approx 1.62$ in $d=2$~\cite{colaiori}. Despite this controversy,
various works~\cite{chin,marinari,shim} confirm the power-counting property
for models with restricted height differences (RSOS, BCSOS) and suggest that
the skewness
\begin{equation}
S \equiv {{W_3}\over{{W_2}^{3/2}}}
\label{defskew}
\end{equation}
and the kurtosis
\begin{equation}
Q \equiv {{W_4}\over{{W_2}^{2}}}-3
\label{defkurt}
\end{equation}
are universal at the steady state.
According to Chin and den Nijs, these universal values do not depend on the
sign of the coefficient $\lambda$ of the nonlinear term of the KPZ equation
(\ref{kpz}).

Consequently, the value of KPZ exponents in $d=2$ is still an open question,
and the universality of the values of the skewness and
the kurtosis deserves to be tested in models other than those with restricted
height differences. The main contribution of this paper is the
analysis of simulations results of a recently proposed model for etching of
crystalline solids~\cite{mello} which belongs to the KPZ class. 
Additional support to some of our conclusions will be provided by simulations
results of ballistic deposition and of the temperature-dependent BCSOS
model~\cite{chin}. For the etching model, we will obtain $\alpha \approx 0.38$
after a detailed analysis of finite-size corrections. Characteristic
relaxation times will be calculated independently and will provide estimates of
the dynamical exponent $z>1.6$, while estimates of the growth exponent give
$\beta \approx 0.23$. Although our error bars intercept those obtained from
extensive simulations of the RSOS model~\cite{marinari}, our central estimates
of $\alpha$ are smaller and, consequently, more distant from the theoretically
proposed value $\alpha=0.4$~\cite{lassig}. On the other hand, our estimates
are near those by Colaiori and Moore~\cite{colaiori} from renormalization
under the mode-coupling approximation. We will also show that our data for
various models confirm the universality of the steady state skewness and
kurtosis. Concerning the steady state skewness, although its absolute value is
universal, its sign changes with the nonlinear coefficient $\lambda$ of the
KPZ equation.

This paper is organized as follows. In Sec. II we briefly describe the
models, the simulation procedure, the methods to estimate $W_n$ and the method
to calculate characteristic relaxation times. In Sec. III, we analyze the
skewness and the kurtosis at the steady states. In Sec. IV, we analyze the
finite-size estimates of the scaling exponents of the etching model, also
showing some results for BD. In Sec. V we summarize our results and present our
conclusions.

\section{Models and simulation procedure}
\label{secmodel}

The model for etching of a crystalline solid of Mello et al~\cite{mello} is
illustrated in Fig. 1 in its growth version. The solids
have square and simple cubic lattice structures in $1+1$ and $2+1$ dimensions,
respectively. At each growth attempt, a column $i$, with current
height $h(i)\equiv h_0$, is randomly chosen. Then its height is
increased by one unit ($h(i)\rightarrow h_0+1$) and any neighboring
column whose height is smaller than $h_0$ grows until its height becomes $h_0$.
One time unit corresponds to $L^2$ growth attempts in $2+1$ dimensions. In the
true etching version of this model, the columns' heights decrease by the same
quantities above. However, in this paper we will always refer to the growth
version of Fig. 1 as "the etching model".

In the ballistic deposition (BD) model, particles are
sequentially released from randomly chosen positions above the substrate,
follow a trajectory perpendicular to the surface and stick upon first
contact with a nearest neighbor occupied site~\cite{barabasi,vold}.

In the BCSOS model defined by Chin and den
Nijs~\cite{chin}, the substrate is a square lattice and the heights $h$ in the
first (second) sublattice are restricted to assume even (odd) values.
Also, the nearest neighbor columns always differ in height by $\Delta
h=\pm 1$. The energy of a given heights configuration $\{ h\}$ is given by
$E\left( \{ h\} \right) = \sum_{\left< i,j\right>}{{1\over 4}K {\left(
h_i-h_j\right)}^2}$, where $K$ is an inverse temperature parameter
and the sum runs over all next nearest neighbor pairs. At each deposition
attempt, a column $c$ is randomly chosen and, if
the constraint of the heights difference is satisfied, then $h(c)\to h(c)+2$
with probability $p\equiv min\left( 1,exp\left( -\Delta E\right) \right)$,
where $\Delta E$ is the energy change if the deposition takes place. The
corresponding coefficient $\lambda$ (Eq. \ref{kpz}) changes sign at a critical
point $K_c$~\cite{chin,amar1}. 

The etching model was simulated in lattices of lengths $L=2^n$ ($n=5$ to
$n=10$) and lattices of lengths $L=2^m\times 50$ ($m=0$ to $m=4$). The
maximum deposition time ranged from ${10}^3$ for the smallest lattices to
$6\times {10}^4$ for the largest ones. During half of this time or more, the
systems have undoubtedly attained their steady states. For the smallest
lattices, $2\times{10}^4$ realizations were simulated, and nearly $300$
realizations for the largest lattices ($L=800$ and $L=1024$).
BD was simulated in lattices of lengths $L=2^n$, from $n=5$
($L=32$) to $n=11$ ($L=2048$). For the analysis of the skewness and of the
kurtosis, it was essential to simulate this model in very large lattices,
thus the number of realizations was relatively small: typically ${10}^3$
realizations for $L\leq 512$, $35$ for $L=1024$ and $8$ for $L=2048$. 
The temperature-dependent BCSOS model was simulated with $K=0.25$ ($L=16$ to
$L=128$) and $K=1.0$ ($L=16$ to $L=512$). For $L\leq 256$,
${10}^3$ different realizations were considered, and $40$ realizations for
$L=512$.

The procedure to calculate average quantities, described below, followed the
same lines for all models. It was previously used
in the analysis of other growth models in $1+1$ and $2+1$
dimensions~\cite{balfdaar,tau,dtwv}.

One important point is the criterion to determine the initial time $t_{min}$
for estimating average quantities at the steady state. In this regime, the
interface width $\xi$ fluctuates around
an average value instead of increasing systematically, which was the case in
the growth and in the crossover regions. Thus, for a fixed lattice length $L$,
the first step was choosing a time interval $t_{min}\leq t\leq t_{max}$, with
nearly constant $\xi$, from visual inspection of the $\xi\times t$ plot
($t_{max}$ was always the maximum simulation time). Subsequently, two tests
were performed. For the first test, the interval was divided in
$5$ subintervals and the average value of $\xi$ was calculated in each one,
forming a sequence of estimates $\{\xi(i)\}$, with $i=1,\dots 5$. If
$\xi(i)<\xi(i-1)$ at least two times along this sequence, then the average
value of $\xi$ in the region $t_{min}\leq t\leq t_{max}$,
$\xi_{sat}^{trial}$, was calculated. In the second
test, from the plot of $\log{(\xi_{sat}^{trial}-\xi(t))}\times t$, for $t$ in
the crossover region, we obtained a rough estimate of the characteristic time
$\tau$ of relaxation to the steady state, as shown in Ref. \protect\cite{tau}.
If $t_{min}>10\tau$, then the interval $t_{min}\leq t\leq t_{max}$ was
accepted as representative of the steady state. Otherwise, a larger
value of $t_{min}$ was chosen and the tests were repeated (it seldom occurred).

In order to estimate the moments of the height distribution and their error
bars, we used their average values within the five subintervals defined above.
Final estimates of $W_n$ are averages of these values, and error bars
were obtained from their variances. 

The dynamical exponent was estimated using a recently proposed
method~\cite{tau}, in which a characteristic time $\tau_0$, proportional
to the relaxation time $\tau$, is calculated. For fixed $L$, after
calculating the saturation width $\xi_{sat}(L)$, $\tau_0$ is defined through
\begin{equation}
\xi{\left( L,\tau_0\right)} = k\xi_{sat}{\left( L\right)} ,
\label{deftau0}
\end{equation}
with a constant $k\lesssim 1$. From relation (\ref{fv}), we
obtain
\begin{equation}
\tau_0\sim L^z .
\label{scalingtau0}
\end{equation}
Typically, the uncertainty in $\tau_0$ is much smaller than that of
$\tau$, estimated from $\log{(\xi_{sat}-\xi)}\times t$ plots - see Ref.
\protect\cite{tau}.

We estimated $\tau_0$ with $k$ ranging from $k=0.5$ to $k=0.8$. Although
these constants are not much different, the values of $\tau_0$ for $k=0.5$ and
$k=0.8$ typically differ by a factor $4$, for fixed $L$. This method was
already applied with success to calculate the dynamical exponent of other
models in various universality classes~\cite{tau,dtwv}.

\section{Universality of skewness and kurtosis}

In Fig. 2a we show the steady state skewness $S\left( L,t\to\infty\right)$ of
the etching model as a function of the inverse lattice length, which provides a
good linear fit of the data, with an asymptotic
estimate $S=0.26\pm 0.01$. The absolute value of $S$ agrees with results of
Chin and den Nijs~\cite{chin}, Shim and Landau~\cite{shim} and Marinari et
al~\cite{marinari} for the RSOS model of Kim and Kosterlitz~\cite{kk} and for
BCSOS models, but those authors obtained $S$ with negative sign. Here, the
positive sign is related to the presence of sharp hills (see process at the
left in Fig. 1) and wide valleys at the surface of the deposit, the opposite
being observed in RSOS deposits.

Note that Eq. (\ref{kpz}) is invariant under the transformations $h\to -h$ and
$\lambda\to -\lambda$, without changing the other parameters. This
transformation changes the sign of the skewness. Consequently, we expect that
the sign of $S$ is related to that of $\lambda$, as previously
observed in the growth regimes of $1+1$-dimensional KPZ
systems~\cite{krughhmeakin}. In fact, in the true etching version of this
model, with erosion leading to decreasing heights, the sign of $S$ changes,
corresponding to the transformation $h\to -h$, $\lambda\to -\lambda$.

Results for the other models contribute to this discussion.
In Fig. 2b we show $S\left( L,t\to\infty\right)$ versus $1/L$ for the BD
model. Note that $S$ is negative for small lattices
(typically $L<500$), but for large lattices it becomes positive, showing that
there are significant morphological differences between the steady states of
small lattices and those of very large systems. This is the main reason to
avoid extrapolating the data in Fig. 2b, even choosing extrapolation variables
other than that $1/L$ (this abscissa in Fig. 2b was chosen only for
illustrate the evolution of $S$ with $L$). Other consequences of this complex
finite-size behavior were previously discussed in Ref.
\protect\cite{balfdaar}. However, it is clear from Fig. 2b that $S$ is
asymptotically positive.

In Figs. 2c and 2d we show the steady state skewness for the
temperature-dependent BCSOS models with $K=0.25$ and $K=1.0$, respectively.
For $K=0.25$, the extrapolation of the three last data points give $S =
-0.28\pm 0.015$, reinforcing the conclusion on the universality of $|S|$.
Due to the small number of data points, the asymptotic correction term
may be other than $1/L$, but it would not affect that conclusion. For $K=1.0$,
the skewness is always positive and rapidly increases with $L$, thus
no extrapolation variable of the form $L^{-\Delta}$, with $\Delta>0$, provides
a reasonable linear fit. However, $S$ will certainly converge to a positive
value as $L\to\infty$, showing that the sign of the steady state skewness
changes as $K$ increases.

The sign of the corresponding parameter $\lambda$ is obtained from the size
dependence of the growth velocity. The steady state growth velocity,
$v_s\left( L\right)$, and the velocity in an infinitely large substrate at
long times, $v_\infty$, obey the relation~\cite{krugmeakin}
\begin{equation}
v_s\left( L\right) = v_\infty - a\lambda L^{-\alpha_\|} ,
\label{velocity}
\end{equation}
where $\alpha_\| = 2\left( 1-\alpha\right)$ and $a$ is positive. Considering
$\alpha=0.38$ (see Sec. IV), we obtain $\alpha_\| =
1.24$. In Figs. 3a-3d, we plotted $v_s\left( L\right)$ versus $x\equiv
1/L^{1.24}$ for the four models. For the etching model, the BD model and the
BCSOS model with $K=1.0$ (Figs. 3a,b,d), $v_s$ decreases with $x$, which
gives a positive $\lambda$, while the opposite occurs in the BCSOS model with
$K=0.25$. It confirms that the steady state skewness of the KPZ equation has
the same sign of the parameter $\lambda$, its absolute value being universal.
The same conclusion was derived from the time dependence of the
velocity in the growth regime~\cite{krugmeakin}.

At this point, we recall that Chin and den Nijs~\cite{chin}
simulated the temperature-dependent BCSOS model with $K=-0.25$ and
$K=0.25$, obtaining $S\approx -0.26$ in both cases. They concluded that $S$ did
not depend on $\lambda$, but they did not estimate the values of this
parameter. Since $\lambda$ is expected to cross zero for a positive $K$, we
conclude that their work failed to consider the regime of positive $\lambda$,
which is represented here by $K=1.0$.

Now we turn to the analysis of the kurtosis (Eq. \ref{defkurt}).
In Fig. 4a we show the steady state $Q$ for the etching model, as a
function of $1/L$. The size dependence is much weaker than that of the
skewness, so that extrapolation variables other than $1/L$ do not have a
significant influence on the asymptotic estimate, $Q=0.134\pm 0.015$
(the large error bar is mainly a consequence of the uncertainties of the
finite-size data). This estimate agrees with previous ones for models
with restricted heights differences~\cite{chin,shim,marinari}, suggesting that
the steady state kurtosis is also universal. In Fig. 4b, we show the data for the BD model and for the BCSOS model with $K=0.25$ and $K=1.0$, which have significant finite-size dependence (in particular those for BD). Thus, the extrapolated values have very large error bars, but are still consistent with a universal value of $Q$.

\section{Roughness, dynamical and growth exponents}

Our first step to estimate the roughness exponent was to
calculate effective exponents $\alpha_{\left( L,i\right)}$ defined as
\begin{equation}
\alpha_{\left( L,i\right)} \equiv { \ln \left[\xi_{sat}\left( L\right)
/\xi_{sat}\left( L/i\right)\right] \over \ln{i} } .
\label{defalphaL}
\end{equation}
It is expected that $\alpha_{\left( L,i\right)}\to \alpha$ for any value of
$i$.

Using different values of $i$ in Eq. (\ref{defalphaL}), we noticed that
$\alpha_{\left( L,i\right)}$ varied with $L$ typically in the range $0.33\leq
\alpha_{\left( L,i\right)} \leq 0.38$ for $50\leq L\leq 1024$, which suggests
that corrections to the scaling relation (\ref{defalpha}) are relevant. Our
first proposal is to assume the main scaling correction as
\begin{equation}
\xi_{sat}\sim L^{\alpha} (a_0+a_1 L^{-\Delta}) ,
\label{corrections}
\end{equation}
where $a_0$ and $a_1$ are constants. Consequently, $\alpha_{\left(
L,i\right)}$ is expected to vary as
\begin{equation}
\alpha_{\left( L,i\right)} \approx \alpha + B L^{-\Delta} ,
\label{alphaL}
\end{equation}
where
\begin{equation}
B = \frac{(1-i^\Delta)}{\ln(i)}\frac{a_1}{a_0} .
\label{B}
\end{equation}

Our data with $50\leq L\leq 1024$ were analyzed using four values of $i$
in Eq. \ref{defalphaL}: $i=2$, $i=2.56$, $i=3.125$ and $i=4$ (for
noninteger $i$, only three or four effective exponents can be
calculated). For each $i$, we plotted $\alpha_{\left(
L,i\right)}\times L^{-\Delta}$ using several exponents $\Delta$ and
did least squares fits of those plots, from which the linear correlation
coefficients $r(\Delta,i)$ were obtained. Since there is no argument to predict
the value of $\Delta$, we adopted the condition of maximizing the
coefficient $r$ to extrapolate our data. In Table I, we show the
exponents $\Delta$ which gave the largest $r$ (best linear fits)
for each $i$. The procedure is illustrated in Figs. 5a and 5b, in which we
show $\alpha_{\left( L,2\right)}\times L^{-0.55}$ and $\alpha_{\left(
L,2\right)}\times L^{-0.65}$, respectively, with the corresponding linear
fits. The values of $\Delta$ in Figs. 5a and 5b are those which give the best
fits with $i=2$ and $i=4$, respectively (see Table I).

$\Delta$ is expected to be independent of the particular choice of $i$ in Eq.
(\ref{defalphaL}), so the differences between the estimates in Table I are
effects of the maximization of $r$. Moreover, other exponents $\Delta$ near
the values shown in Table I also provided reasonable linear fits of
$\alpha_{\left( L,i\right)}\times L^{-\Delta}$ plots. In other words, large
linear correlation coefficients were also obtained by considering
$0.3\lesssim \Delta\lesssim 0.8$ for different choices of $i$. Consequently, it
is not possible to obtain a reliable estimate of that correction exponent. On
the other hand, the asymptotic $\alpha$ obtained from the same fits fluctuate
within a narrow range.  Accounting for the error bars of the data, we obtained
$\alpha=0.385\pm 0.01$ for $i=2$ and $\alpha=0.382\pm 0.01$ for $i=4$.

We also checked the effect of considering a fixed
correction exponent $\Delta=0.55$ (which is near the values in Table I) for
all values of $i$. This procedure will be called fixed $\Delta$
method. Least squares fits of the $\alpha_{\left(
L,i\right)}\times L^{-0.55}$ plots were performed, providing the slopes $B$
which are shown in Table I. From Eq. (\ref{B}), it is expected that
${{B\ln{i}}\over{(1-i^\Delta)}} = \frac{a_1}{a_0} = const$ (see also Eqs.
\ref{corrections} and \ref{alphaL}), thus we also showed in Table I the
corresponding estimates of ${{B\ln{i}}\over{(1-i^\Delta)}}$. This quantity
fluctuates with $i$, indicating that there is no systematic trend in our
results due to choice of different values of $i$ for calculating
effective exponents. Also notice that the estimates of $\alpha$ from the fixed
$\Delta$ method are in same range of the those obtained from the maximization
of correlation coefficients.

Since the range of lattice lengths considered here is not very large and the
correction exponents $\Delta$ estimated above are relatively small, we
tried to improve our analysis with a different assumption for the
scaling corrections. Contrary to the previous procedure, now we will
consider a well defined form for the main scaling correction, which is an
additional constant term $\xi_I^2$ in the dynamic scaling relation (\ref{fv})
for the squared interface width:
\begin{equation}
\xi^2 \left( L,t\right) = \xi_I^2 + L^{2\alpha} g\left(
tL^{-z}\right) ,
\label{intrinsic}
\end{equation}
where $g$ is a scaling function. $\xi_I$ is called intrinsic width and
represents contributions of small length scales fluctuations, typical of models
with large local heights differences~\cite{wolf,kertesz,moro} such as the
etching model and the BD model.
From Eqs. (\ref{intrinsic}) and (\ref{corrections}), the assumption of the
intrinsic width as the most relevant subleading correction corresponds to a
(fixed) correction exponent $\Delta=2\alpha\approx 0.8$. It is slightly larger
than the typical values obtained in the previous analysis.

Effective exponents $\alpha_L^{\left( I\right)}$ which cancel the
contribution of $\xi_I^2$ are defined as
\begin{equation}
\alpha_L^{\left( I\right)} \equiv {1\over 2}
{ \ln{ \left[ \xi_{sat}^2\left( 2L\right) - \xi_{sat}^2\left( L\right)
\right] /\left[ \xi_{sat}^2\left( L\right) - \xi_{sat}^2\left( L/2\right)
\right] } \over\ln{2} } .
\label{defalphai}
\end{equation}

In Fig. 6a we show $\alpha_L^{\left( I\right)}$ versus $1/L$ for the etching
model and a least squares fit of these data, which provides $\alpha=0.383$
asymptotically. Here, the variable $1/L$ in the abscissa was chosen only to
illustrate the behavior of the $\alpha_L^{\left( I\right)}$ data. It represents
a second correction term for the $\xi$ scaling, which is still more difficult
to measure than the first correction term. Thus, we also tested other variables
in the form $L^{-\Delta_1}$ to extrapolate $\alpha_L^{\left( I\right)}$, with
$0.5\leq \Delta_1\leq 2$ ($\Delta_1 = 1$ was used in Fig. 6a). The
corresponding linear fits give $0.380<\alpha<0.387$. The small range of the
asymptotic $\alpha$ is a consequence of the slow variation of $\alpha_L^{\left(
I\right)}$ with $L$ ($2\%$ from $L=100$ to $L=500$), as shown in Figs. 6a.
Accounting for the error bars of the data, our final estimate is $\alpha =
0.383\pm 0.008$.

The forms of finite-size corrections analyzed above, Eqs. (\ref{alphaL}) and
(\ref{intrinsic}), cannot be rigorously justified, but are based on heuristic
arguments. See e. g. Ref. \protect\cite{barabasi} and references therein.
Certainly, the fact that $\alpha_L^{\left( I\right)}$ increases slowly with $L$
is a support to the assumption that the intrinsic width is the most
relevant correction term to $\xi^2$. Anyway, using different assumptions on
the scaling corrections was essential to confirm the reliability of the above
estimate of the roughness exponent.

The universality of the values of skewness and kurtosis imply that $W_3$ and
$W_4$ may also be used to estimate $\alpha$. The effective
exponents obtained from the third moment have very large error bars, but those
obtained from the fourth moment behave similarly to the ones obtained from the
interface width. The analogous of exponents $\alpha_{\left( L,2\right)}$ and
$\alpha_{\left( L,4\right)}$ (Eq. \ref{defalphaL}) calculated with $W_4$ also
converge to the range $0.38\leq\alpha\leq 0.385$ with strong corrections
to scaling. Effective exponents which cancel the contribution of a constant
additive term in the dynamic scaling relation for $W_4$ (analogous to
$\alpha_L^{\left( I\right)}$) are defined as
\begin{equation}
\alpha_L^{\left( I,4\right)} \equiv
{1\over 4} { \ln{ \left[ W_4\left( 2L,t\to\infty\right) - W_4\left(
L,t\to\infty \right) \right] /\left[ W_4\left( L,t\to\infty\right) - W_4\left(
L/2,t\to\infty\right) \right] } \over\ln{2} } .
\label{defalphai4}
\end{equation}
They are plotted in Fig. 6b as a function of $1/L$. The asymptotic estimate,
obtained with the procedure described above, is $\alpha = 0.379\pm 0.012$.
It is in good agreement with the estimate from the interface width and also
excludes $\alpha = 0.4$.

The same analysis was also performed with the BD model, as shown in
Fig. 7, with data for $L\leq 512$. Although the results are less accurate than
those for the etching model, they also suggest that $\alpha<0.4$
asymptotically.

In order to estimate the dynamical exponent $z$, we
calculated effective exponents $z_L$ defined as
\begin{equation}
z_L \equiv { \ln \left[\tau_0\left( 2L\right)
/\tau_0\left( L/2\right)\right] \over \ln{4} } ,
\label{defzL}
\end{equation}
using the characteristic times $\tau_0$ defined in Sec. II (Eqs. \ref{deftau0} and
\ref{scalingtau0}).
In Figs. 8a and 8b we show $z_L$ versus $1/L$ for the etching model, obtained
using $k=0.6$ and $k=0.8$ to calculate $\tau_0$, respectively. Here,
the abscissa $1/L$ is also chosen to illustrate the evolution of $z_L$,
but not to perform extrapolations of the data. Although the error bar of the
data for $L=800$ is relatively large, those plots indicate that $z>1.6$.
Considering the trend for large $L$ and different values of $k$,
we estimate $1.605\leq z\leq 1.64$. These values are consistent with the above
estimates of $\alpha$ and the exact relation $\alpha +z=2$.

Finally, we estimated the
growth exponent (Eq. \ref{defbeta}) of the etching model with the same
procedure previously applied with success to BD and to the Das Sarma and
Tamborenea model in $1+1$ dimensions~\cite{balfdaar,dtwv}. The growth
region for each $L$ begins at $t_0 = 50$ and ends at the maximum time
$\tau_{max}$ such that the linear correlation coefficient of the data in the
range $t_0\le t\le\tau_{max}$ exceeds a fixed value $r_{min}$~\cite{balfdaar}.
Here, $r_{min}=0.99995$ and $r_{min}=0.9999$ are considered.
Effective exponents $\beta_L$ are defined as the slopes of the
linear fits of $\log{W}\times\log{t}$ plots using all data in the
above-defined growth regions. In Fig. 9 we plot $\beta_L$ versus $1/L$ for
the etching model using $r_{min}=0.99995$ and $r_{min}=0.9999$. The error bars
of those effective exponents is very small. We also show in Fig. 9 the linear
fits of the $\beta_L\times 1/L$ data for each $r_{min}$. Other variables in
the form $L^{-\Delta}$ were used to extrapolate the $\beta_L$ data, giving
asymptotic estimates $\beta = 0.229\pm 0.005$. Within error
bars, it agrees with the value $\alpha/z = 0.234\pm 0.009$ obtained from the
above estimates of $\alpha$ and $z$.

\section{Conclusion}

We studied four $2+1$ dimensional discrete growth models in the KPZ class,
determining critical exponents and steady state values of the skewness $S$ and
the kurtosis $Q$. Accurate estimates of the scaling exponents were obtained for
the etching model proposed by Mello et al~\cite{mello}: $\alpha = 0.383\pm
0.008$, $1.605\leq z\leq 1.64$, $\beta = 0.229\pm 0.005$. Results for the
ballistic deposition model also indicate that $\alpha<0.4$. The
presence of the intrinsic width as the main correction to the
interface width scaling was considered to extrapolate the simulations data. We
also obtain the estimates $S=0.26\pm 0.01$ and $Q=0.134\pm 0.015$ in the
steady state regime of the etching model. Results for the BD
model and of the BCSOS model, together with previous results for the RSOS model, suggest that the absolute value of $S$ and the
value of $Q$ are universal, the sign of the skewness being
the same of the parameter $\lambda$ of the corresponding KPZ equation.

The above estimate intercepts the error bar of the roughness exponent
of the RSOS model given by Marinari and co-workers~\cite{marinari}, $\alpha =
0.393\pm 0.003$. However, our central estimate is significantly lower than
theirs, and the result for BD confirm this trend. On the other hand, our
estimate is very near that by Colaiori and Moore, $\alpha\approx 0.38$, from
renormalization methods~\cite{colaiori}. Additional support to our conclusions
was given by the independent calculation of exponents $z$ and $\beta$,
contrary to recent simulation works on these lines~\cite{chin,marinari}, which
were limited to the calculation of exponent $\alpha$.

We believe that much more accurate estimates of $\alpha$ are difficult to be
achieved with numerical simulations of this type of lattice model. However,
we consider that this work provides a significant amount of numerical results
indicating that the theoretically proposed value $\alpha = 0.4$~\cite{lassig}
is not valid, within the limits of the assumptions made about the form of finite-size scaling corrections. This result and the evidence of universality of the values of the skewness and the kurtosis may motivate further analytical (maybe also numerical) studies of the KPZ theory in $2+1$ dimensions.


\begin{table}
\caption{For each $i$ in Eq. (\ref{defalphaL}), are given: the values of the
correction exponent $\Delta$ which provided the largest correlation
coefficients $r$ of fits of $\alpha_{\left( L,i\right)}\times L^{-\Delta}$
and the corresponding asymptotic $\alpha$, the values of the slope $B$
of the fits with fixed $\Delta=0.55$ and the corresponding values of
asymptotic $\alpha$ and of $B\ln{i}/(1-i^\Delta)$.}
\vskip 0.2cm
\halign to \hsize
{\hfil#\hfil&\hfil#\hfil&\hfil#\hfil&\hfil#\hfil&\hfil#\hfil&\hfil#\hfil&\hfil#\hfil\cr
$i$ (Eq. \ref{defalphaL}) & \ $\Delta$ & \ $\alpha$ & \ $B$ & \ $\alpha$ & \ $B\ln{i}/(1-i^\Delta)$ \cr
& \ (largest $r$) & \ (largest $r$) & \ ($\Delta
=0.55$ fixed) & \ ($\Delta =0.55$ fixed) & \ ($\Delta =0.55$ fixed) & \cr
$2$ & \ $0.55$ & \ $0.384$ & \ $0.472$ & \ $0.384$ & \ $0.705$ \cr
$2.56$ & \ $0.50$ & \ $0.383$ & \ $0.434$ & \ $0.381$ & \ $0.603$ \cr
$3.125$ & \ $0.6$ & \ $0.383$ & \ $0.553$ & \ $0.385$ & \ $0.723$ \cr
$4$ & \ $0.65$ & \ $0.381$ & \ $0.545$ & \ $0.383$ & \ $0.661$ \cr  }
\label{table1}
\end{table}

\begin{figure}
\caption{The growth rules of the model of Mello et al (original growth
version). The arrows indicate the randomly chosen columns and the new
particles are shown in gray.}
\label{fig1}
\end{figure}

\begin{figure}
\caption{Steady state skewness of various models versus inverse lattice length:
(a) etching model (in the growth version); (b) ballistic deposition; (c) BCSOS
model with $K=0.25$; (d) BCSOS model with $K=1.0$. The solid lines in (a) and
(c) are least squares fits of the data.}
\label{fig2} \end{figure}
 
\begin{figure}
\caption{Steady state velocities versus $1/L^{\alpha_\|}$, with $\alpha_\| =
1.24$, for: (a) etching model; (b) ballistic deposition model; (c) BCSOS model
with $K=0.25$; (d) BCSOS model with $K=1.0$. Error bars are smaller than the
size of the data points.}
\label{fig3}
\end{figure}  
 
\begin{figure}
\caption{Steady state kurtosis of various models versus inverse
lattice length: (a) etching model; (b) ballistic deposition model (squares),
BCSOS model with $K=0.25$ (triangles) and BCSOS model with $K=1.0$
(crosses). Error bars in (b) (not shown) are smaller than the size of the data
points, except for BD in the largest lattice ($L=1024$). }
\label{fig4}
\end{figure}       
                        
\begin{figure}
\caption{Effective exponents $\alpha_{\left( L,2\right)}$ and $\alpha_{\left(
L,4\right)}$ for the etching model versus $L^{-\Delta}$, with the exponents
$\Delta$ that give the best linear fits for $i=2$ and $i=4$, respectively.}
\label{fig5}
\end{figure}

\begin{figure}
\caption{Effective exponents $\alpha_L^{\left( I\right)}$ (from interface
width) and $\alpha_L^{\left( I,4\right)}$ (from the fourth moment of the
height distribution) for the etching model versus inverse lattice length.}
\label{fig6}
\end{figure}

\begin{figure}
\caption{Effective exponents $\alpha_L^{\left( I\right)}$ for the ballistic
deposition model versus inverse lattice length.}
\label{fig7}
\end{figure} 

\begin{figure}
\caption{Effective exponents $z_L$ for the etching model versus inverse lattice
length. Relaxation times $\tau_0$ were obtained with: (a) $k=0.6$ and (b)
$k=0.8$ in Eq. (\ref{deftau0}).}
\label{fig8}
\end{figure} 

\begin{figure}
\caption{Effective exponents $\beta_L$ for the etching model versus inverse
lattice length, obtained with minimum correlation coefficients
$r_{min}=0.9999$ (squares) and $r_{min}=0.99995$ (triangles). Solid lines are
least squares fits of each set of data ($200\leq L\leq 1024$).}
\label{fig9}
\end{figure} 

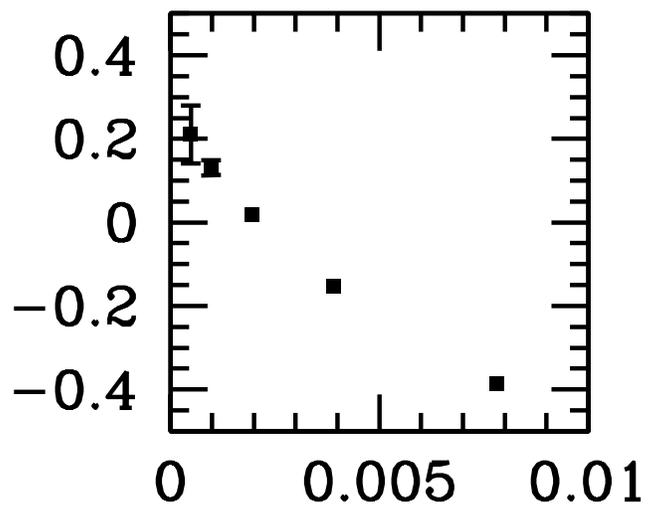
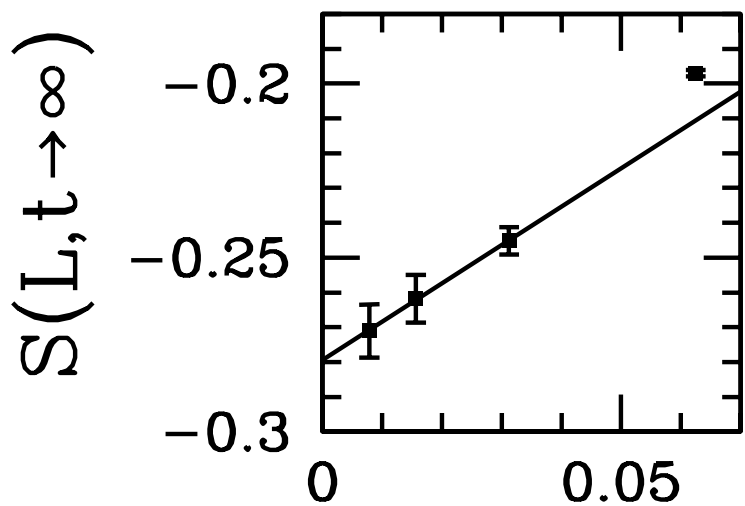
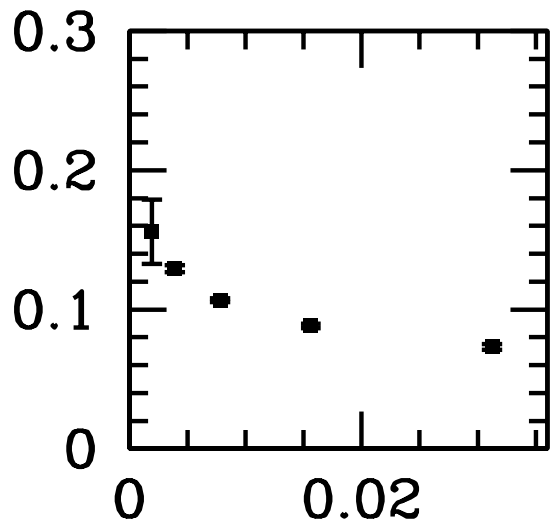
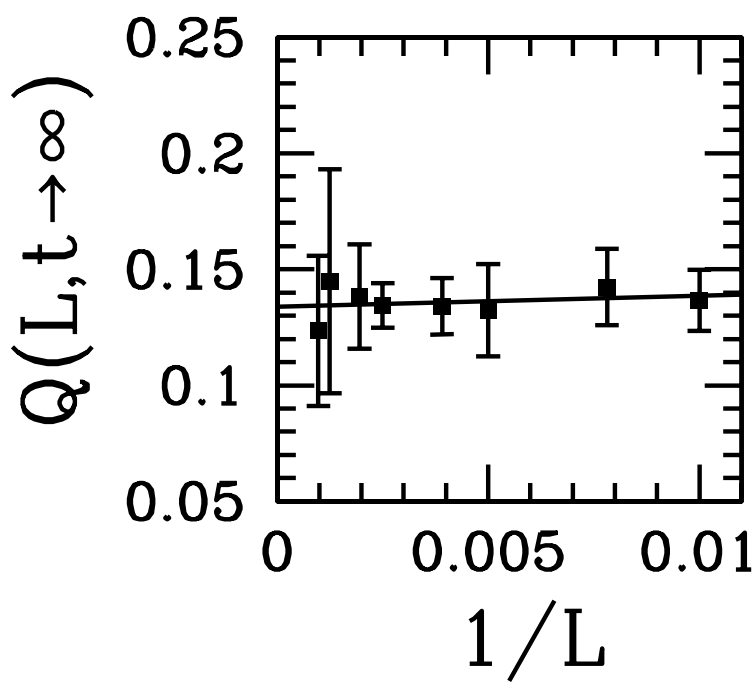
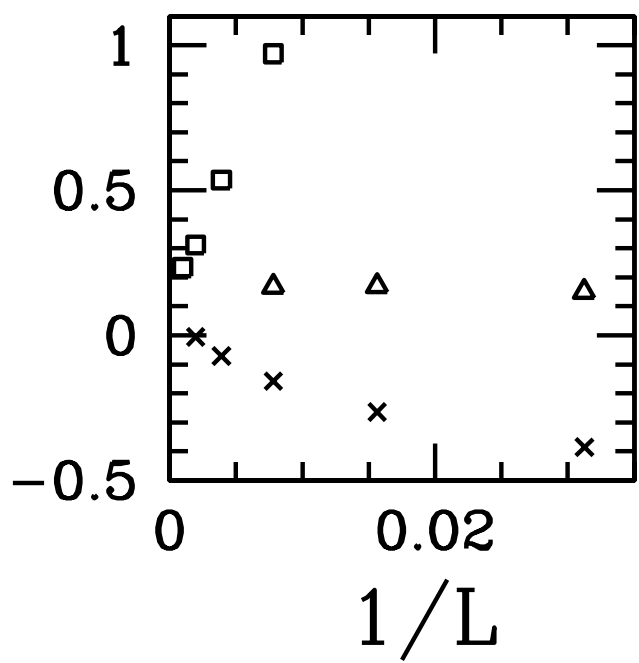
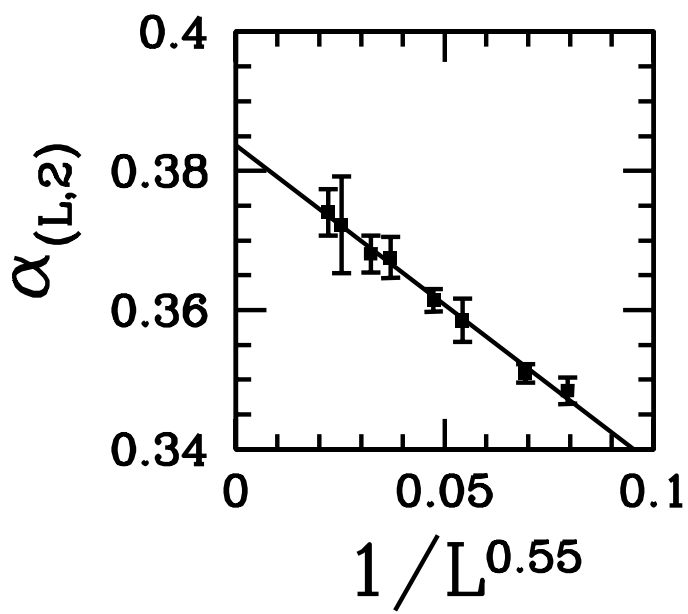
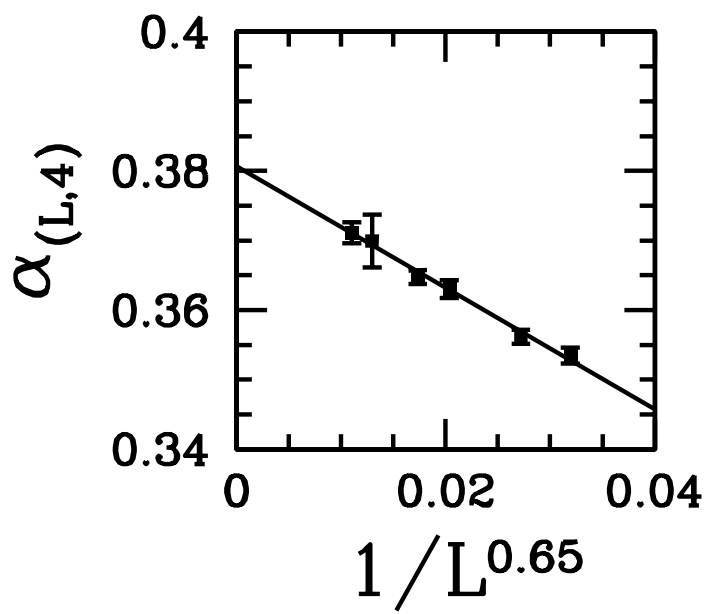
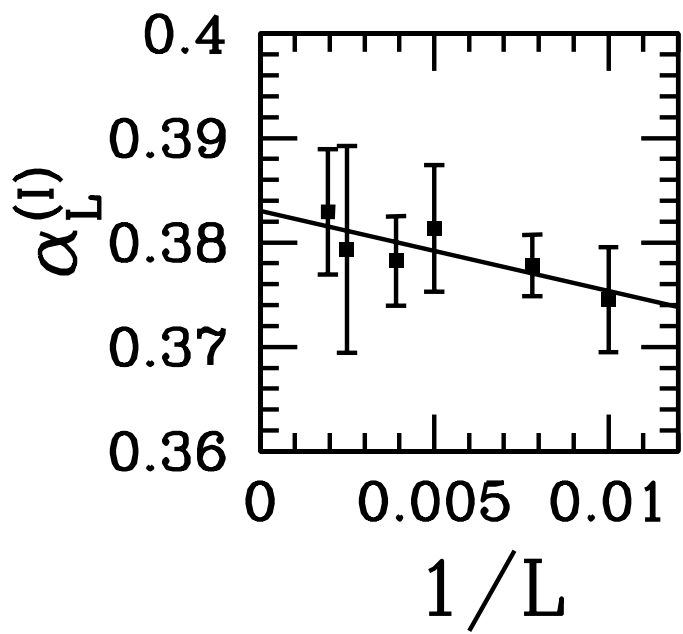
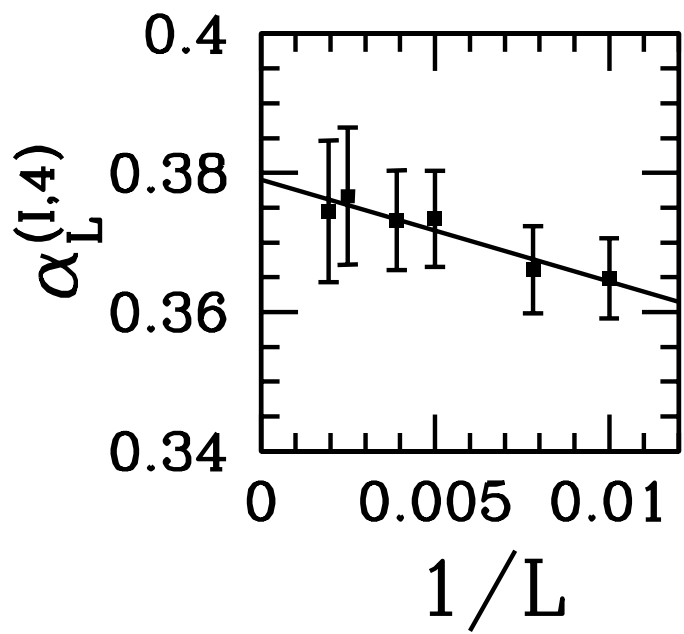
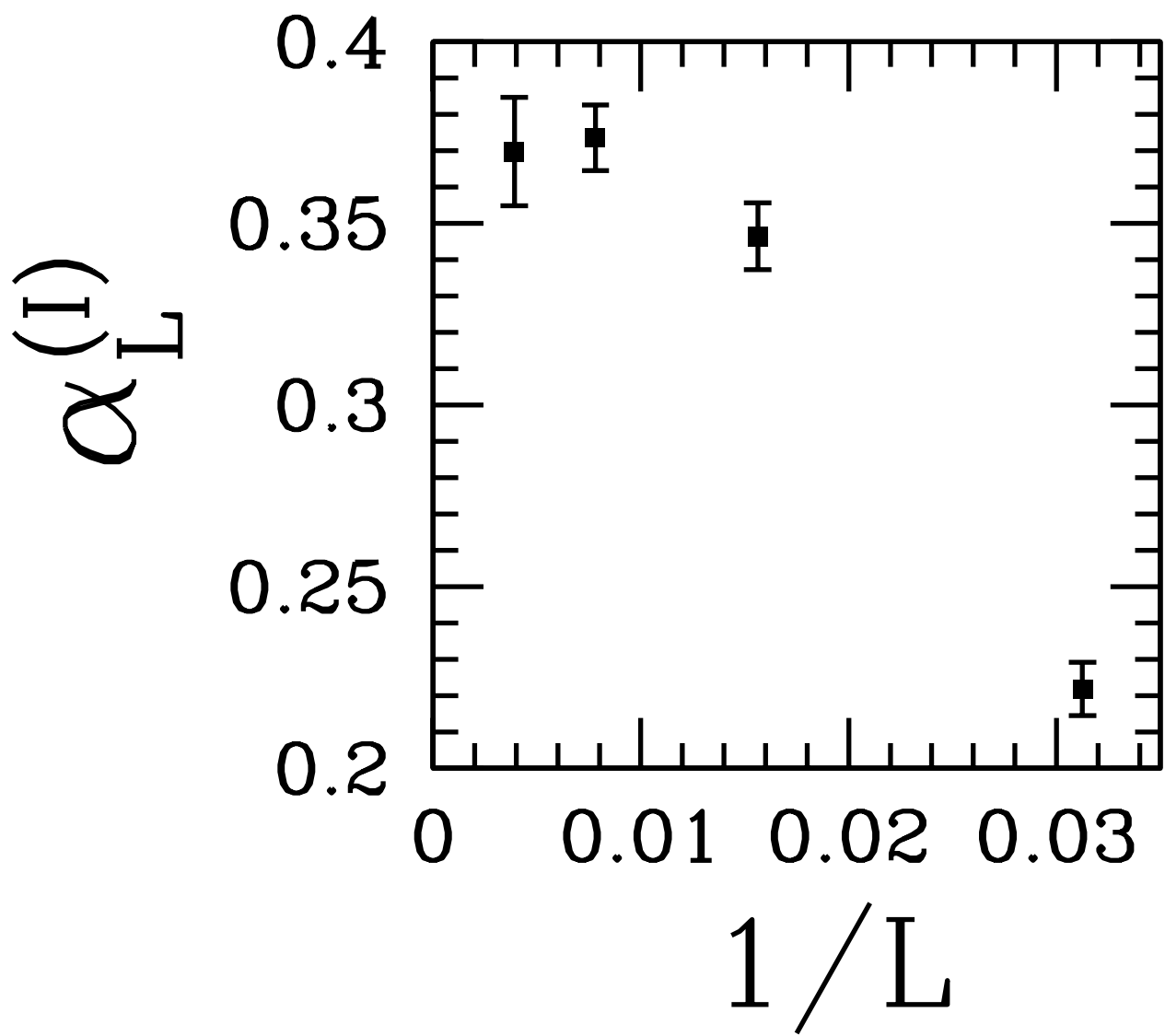
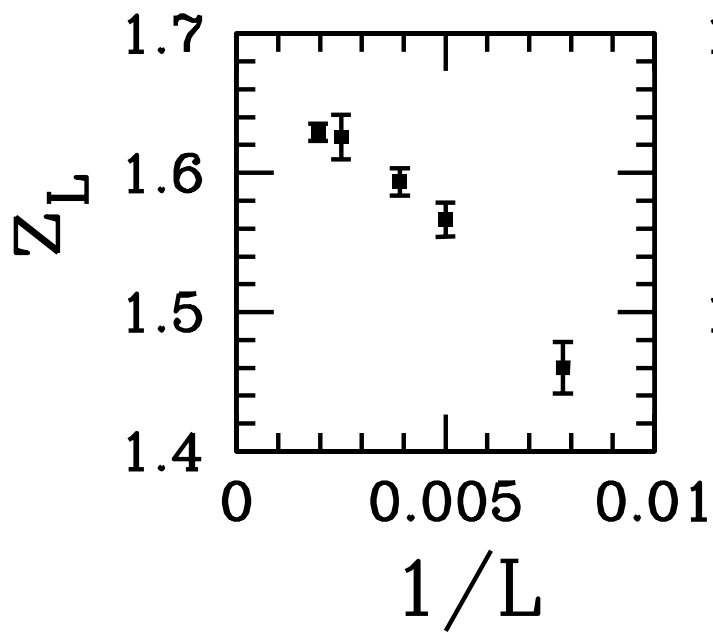
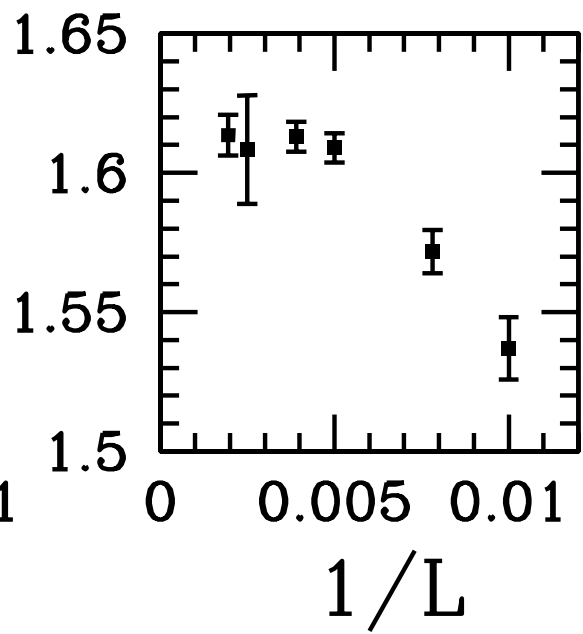
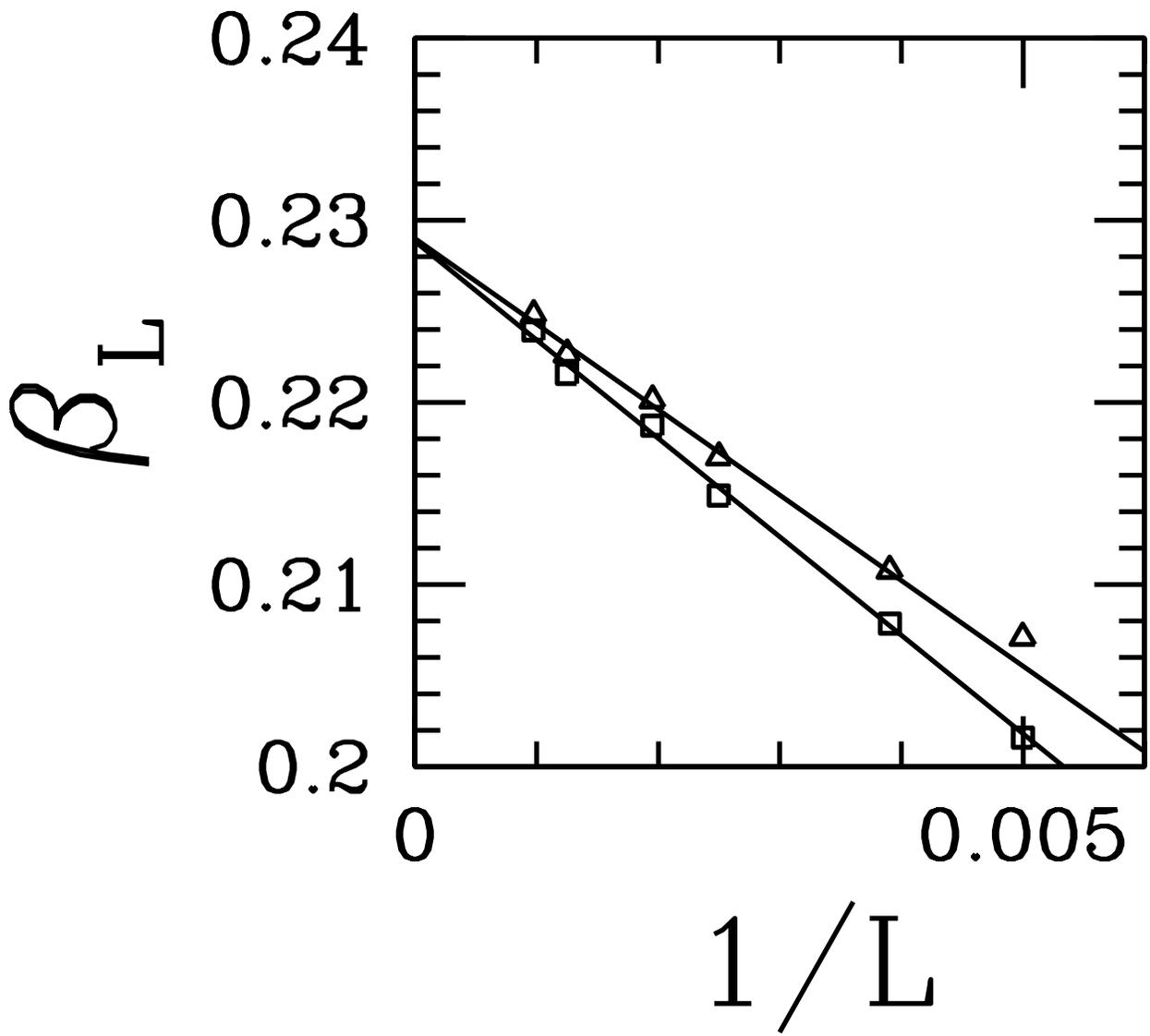
\end{document}